\documentclass{aa}  

\usepackage{graphicx}
\usepackage{txfonts}
\usepackage{amsmath, amssymb}
\usepackage{xcolor}
\usepackage{natbib}
\usepackage[colorlinks=true, linkcolor=blue, citecolor=blue, urlcolor=blue]{hyperref}
\begin{document}

   \title{Pulse frequency variations and timing noise of MXB 0656--072 during the 2007--2008 type I outbursts and implications for its magnetic field}

   \authorrunning{Serim, M. M. et al.}
   \titlerunning{MXB 0656--072}


   \author{M. M. Serim\inst{1}\thanks{E-mail: serim@astro.uni-tuebingen.de},
          D. Serim\inst{1},
          Ç. K. Dönmez\inst{2},
          Y. Tuo\inst{1}\thanks{E-mail: youli.tuo@astro.uni-tuebingen.de},
          L. Ducci\inst{1,3},
          A. Baykal\inst{2}\and
          A. Santangelo\inst{1}
          }

   \institute{Institut fur Astronomie und Astrophysik, Eberhard Karls Universit\"{a}t, Sand 1, D-72076 Tübingen, Germany
         \and
         Department of Physics, Middle East Technical University, 06800 Ankara, Turkey
         \and
         ISDC Data Center for Astrophysics, Universit\'e de Gen\`eve, 16 chemin 
d'\'Ecogia, 1290 Versoix, Switzerland}

   \date{Received XXX; accepted XXX}

 
  \abstract
   {}
   {We aim to explore the properties of the Be/X-ray binary system MXB~0656–072 from a timing analysis perspective through an investigation of the RXTE/PCA and \emph{Fermi}/GBM data during its 2007–2008 type I outbursts.}
   {We applied two new techniques, for the first time, along with the conventional Deeter method to produce higher-resolution power density spectra (PDS) of the torque fluctuations. We also investigated the spin frequency evolution of the source by utilising a pulse timing technique.}
   {The PDSs show a red noise pattern, with a steepness of $\Gamma \sim -2$ and a saturation timescale of $\sim$150 d, indicating that MXB~0656–072 is a disc-fed source. With the obtained long term spin frequency evolution, we  reveal the torque–luminosity correlation of MXB~0656–072 for the first time.  We also demonstrate that the frequency evolution is largely consistent with the Ghosh–Lamb model. In the RXTE/PCA observations, the pulsed emission disappears below $\sim$5$\times 10^{35}$ erg s$^{-1}$, while the profiles remain stable above this value in our analysis time frame. We show that the magnetic field strength deduced from the torque model is compatible with the field strength of the pulsar derived from the cyclotron resonance scattering feature. Utilising the new distance of MXB~0656–072 measured by \emph{Gaia}, we show that the spectral transition of MXB~0656–072 occurs at a luminosity that matches the expected theoretical transition from the subcritical to supercritical accretion regime.}
   {}


   \maketitle
%

\section{Introduction}
\label{sec:1}


MXB~0656--072 is a transient BeXRB system consisting of a neutron star with a spin period of 160.7 s \citep{2003Morgan} and a O9.7Ve spectral type donor star \citep{2003Pakull}.
Despite it being among the earliest transients discovered \citep{1975Clark}, this source still stands as a relatively less explored member.
Following its discovery, MXB~0656--072 entered a state of quiescence, only to rebrighten in 2003 during a two-month outburst episode between October and December.
Analysis of RXTE data by \citep{2006McBride} classified the outburst as type II \citep[see e.g.][for a review about BeXRBs]{2011Reig}, revealing a cyclotron resonance scattering feature (CRSF) at $\sim$33 keV, and implying a magnetic field strength of $\sim$$3.7 \times 10^{12}$ G.
Furthermore, they estimated a distance of 3.9$\pm$0.1 kpc using optical observations.
However, the source is reported to be at a distance of 5.7$\pm$0.5 kpc in the \emph{Gaia} Early Data Release 3 (EDR3) catalogue (Source ID: 3052677318793446016), which is not entirely consistent with the aforementioned value \citep{2016Gaia, 2021Gaia}.

The second (and the most recent) reappearance of MXB~0656--072 lasted for about one year.
The source exhibited a series of X-ray bursts between 2007 November and 2008 November, as observed by \emph{INTEGRAL} \citep{2007Kreykenbohm}, RXTE \citep{2007Pottschmidt}, \emph{Swift} \citep{2007Kennea}, and \emph{Fermi}\footnote{\url{https://gammaray.nsstc.nasa.gov/gbm/science/pulsars/lightcurves/mxb0656.html}}.
During the series of type I outbursts, RXTE/PCA--HEXTE spectra show a CRSF at the luminosity of $\sim$$6.6 \times 10^{36}$ erg s$^{-1}$ (for 3.9 kpc) in the 3--22 keV energy band \citep{2012Yan}.
Through the analysis of \emph{Swift}/BAT and RXTE/ASM data, \cite{2012Yan} deduced an orbital period of $\sim$101.2 d; nevertheless, the specific orbital parameters characterising the system are yet to be established.
Their analysis highlighted that the spectra are best described by an absorbed low-energy cut-off power law, accompanied by a CRSF at $\sim$30 keV and an iron line at $\sim$6.4 keV.
Furthermore, \cite{2012Nespoli} probed the multiwavelength variability of the transient system MXB~0656--072.
In addition to studies on outbursts, MXB~0656--072 has also been studied in its quiescence phase using data from \emph{Chandra} observatory \citep{2017Tsygankov} and, most recently, from \emph{NuSTAR} \citep{2023Raman}.

This paper focuses on the examination of timing and pulse variability of MXB~0656--072 throughout its 2007--2008 outburst.
The structure of the paper is as follows:
In Section \ref{sec2}, the data and the associated reduction tools used in this study are detailed.
Section \ref{sec:timing} outlines the methodology and the results of pulse timing and torque-luminosity analysis, followed by a comprehensive investigation of timing noise behaviour of the source in Section \ref{Sec4}.
Finally, in Section \ref{Sec5}, we discuss the physical interpretations of our results within the context of prior research findings.

\section{Data}\label{sec2}
Our study concentrates on the series of type I outbursts of MXB~0656--072 during 2007--2008.
The outburst sequence was extensively monitored with RXTE/PCA \citep{1996Jahoda} across the observation sets P93032 and P93423.
The RXTE monitoring campaign, started on 2007 November 14 and ended on 2008 November 18, consists of 172 individual observations.
Typical exposures varied between 0.5 and 28 ks throughout these observations.

Analysis of the RXTE/PCA observations are performed using the \textsc{heasoft v6.31} software.
The data filtering included conditions, such as an elevation angle higher than 10 degrees, source offset below 0.02\%, and electron contamination under 0.1.
Additionally, to enhance the signal-to-noise ratio, data collected within 30 minutes of the South Atlantic Anomaly peak were excluded.

Utilising the GoodXenon mode events, we generated a light curve with a 0.125 s time resolution in the 3--20 keV energy band.
Since there are different numbers of proportional counter units (PCUs) active during each observation, we used the \texttt{CORRECTLC} command to adjust the count rates to reflect the simultaneous operation of all five PCUs.
Finally, the photon arrival times of the light curve were corrected to the barycenter for the timing analysis.

 With the aim of measuring the flux level of each observation, we further extracted the spectrum for each observation using only PCU2 top layer data in standard2f mode.
We used the EPOCH 5C model for background generation. Then we used the bright background model, unless the source count rate dropped below <40 cts/s/PCU as suggested by the RXTE team\footnote{\url{https://heasarc.gsfc.nasa.gov/docs/xte/pca_news.html}}.
The overall continuum of the spectra can accurately be modelled by an absorbed cut-off power law, with an iron emission line at 6.4 keV and a CRSF feature around 30 keV \citep{2012Yan,2012Nespoli}. In this study, we focus only on the timing properties of the source. We limit our spectral analysis to the 3--20 keV energy range, within which the pulse timing analysis is conducted, with a sole aim of obtaining the flux evolution of the source to check for any correlation with the frequency derivative and to investigate it in terms of accretion models. Considering the limited energy band, each spectrum is modelled with an absorbed power law to measure the unabsorbed flux levels in the given range. 
We also added a Gaussian line or a cut-off to the model when they are statistically significant.  The resulting 3--20 keV flux measurements were then converted to the luminosity, assuming a distance of 5.7 kpc. For a detailed analysis of the spectral evolution of MXB 0656--072 with the same RXTE data set, we refer  to \cite{2012Yan} and \cite{2012Nespoli}.

\section{Pulse timing and torque--luminosity relation}
\label{sec:timing}
We started by calculating a nominal spin frequency through folding the barycenter-corrected RXTE/PCA light curve at statistically independent trial frequencies \citep{1983Leahy}.
Next, we folded the light curve at the nominal spin period to generate 20 phase-bin pulse profiles for each observation.
In cases where the time difference between consecutive observations was less than 1 day, we merged them together and extracted a combined pulse profile.
To determine the time of arrivals (TOAs) of the pulses, each profile was transformed into a Fourier harmonic representation and cross-correlated \citep{1985Deeter} with the template profile, which was chosen as the one with the highest $\chi^2$ among them.

\begin{figure}
\centering
    \includegraphics[width=0.95\columnwidth]{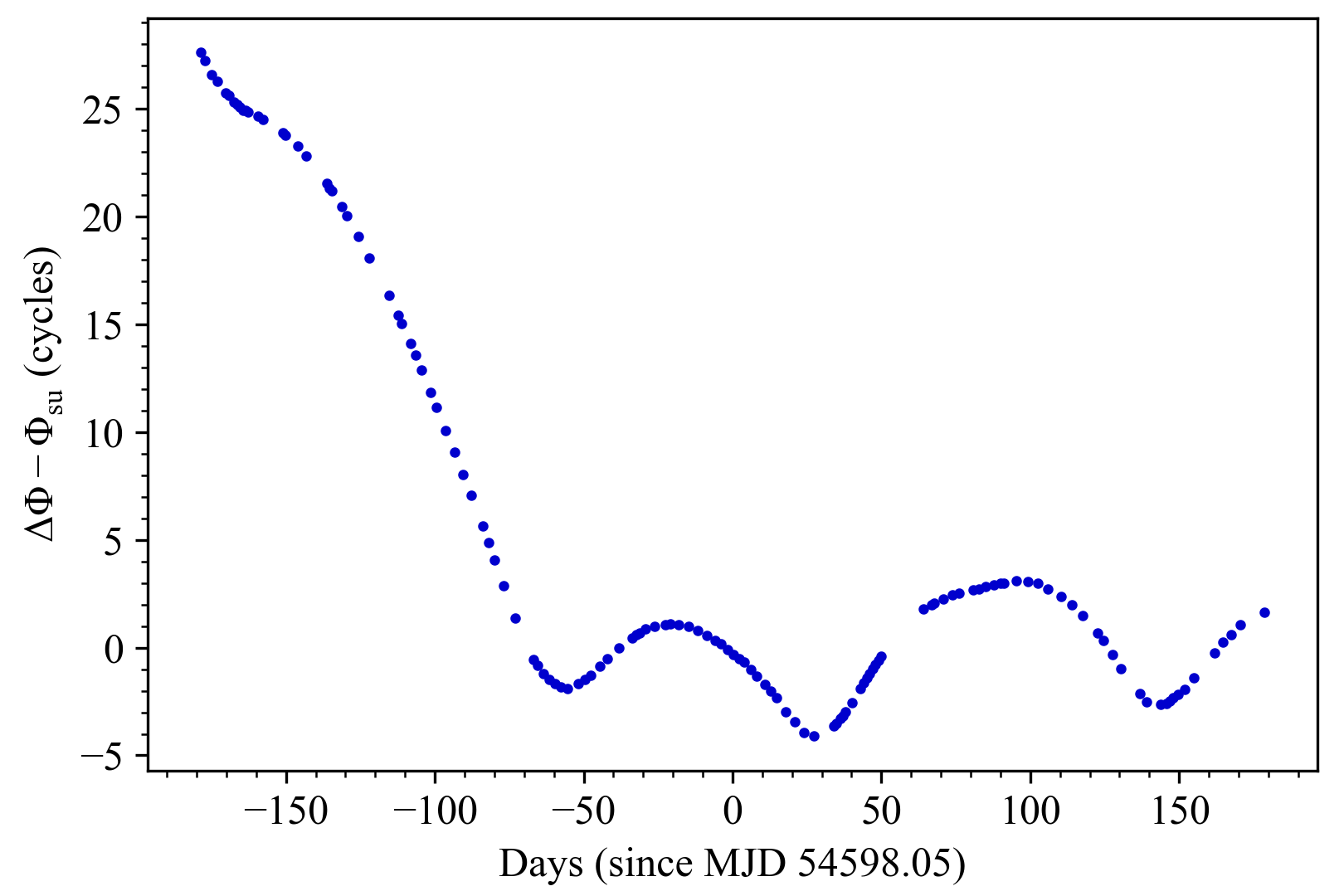}
    \includegraphics[width=0.95\columnwidth]{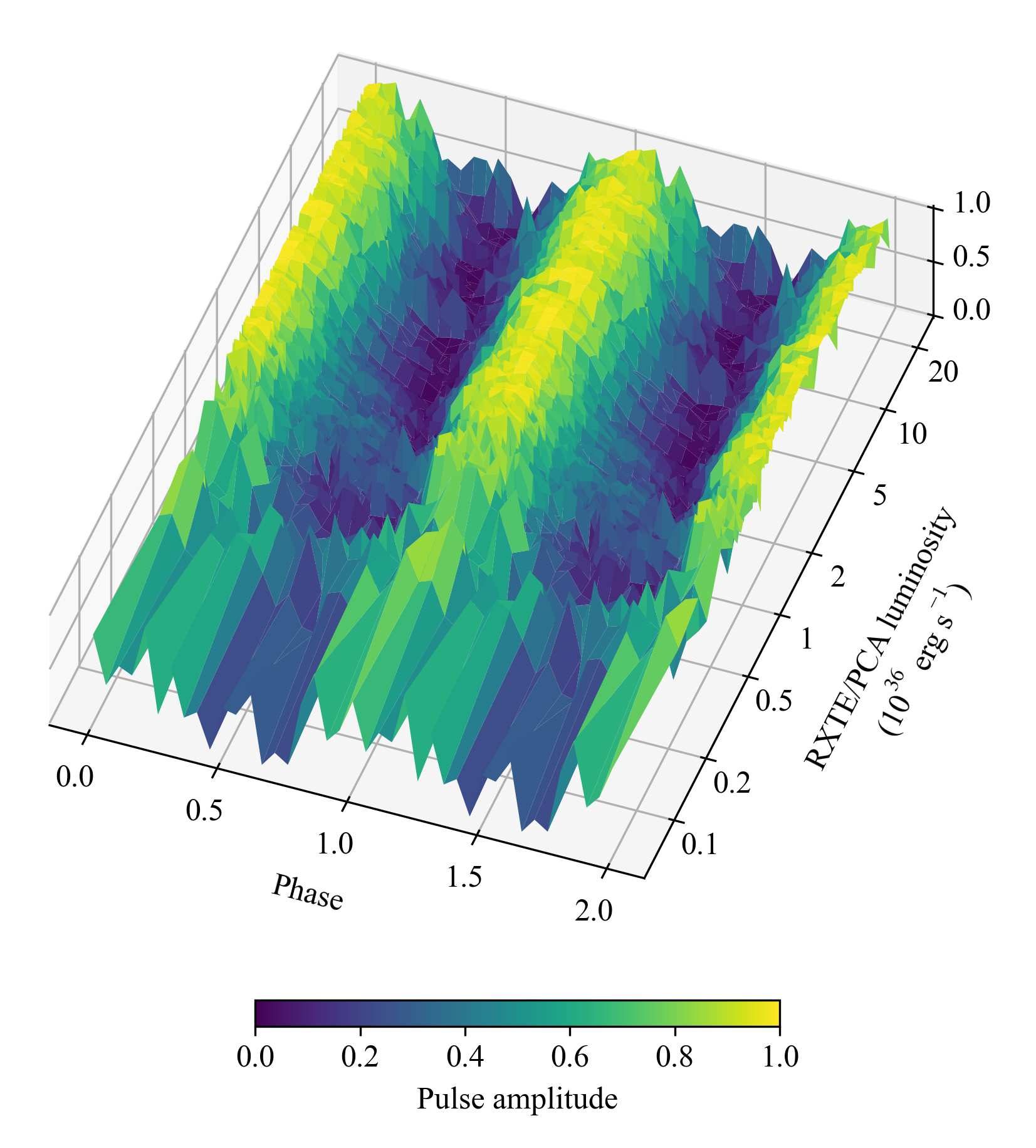}
    \caption{ Pulse TOAs obtained from the RXTE/PCA observations after the removal of the overall spin-up trend (see text), shown at the top. Pulse profile evolution of MXB 0656--072, sorted by 3-20 keV RXTE/PCA luminosity (bottom). To improve legibility, the pulse profiles have been normalised to [0, 1] range and displayed for two complete cycles.}
    \label{fig:f1}
\end{figure}

The observed pulsed emission profiles during this interval possess a single-peaked shape.
We represent the 3D plot of the normalised pulses sorted by RXTE/PCA count rates in Figure \ref{fig:f1}.
The pulse profiles do not exhibit a significant evolution over either luminosity or time.
The only exception is that we cannot observe any pulsed emission in the last few RXTE/PCA observations during which the source luminosity drops below $\sim$$5 \times 10^{35}$ erg s$^{-1}$.
It should be noted that the source emission stays of the order of 10$^{36-37}$ erg s$^{-1}$ within the data set in our analysis and the deduced interpretations were made accordingly.

\begin{figure}
\centering
    \hspace{-4mm}\includegraphics[width=0.96\columnwidth]{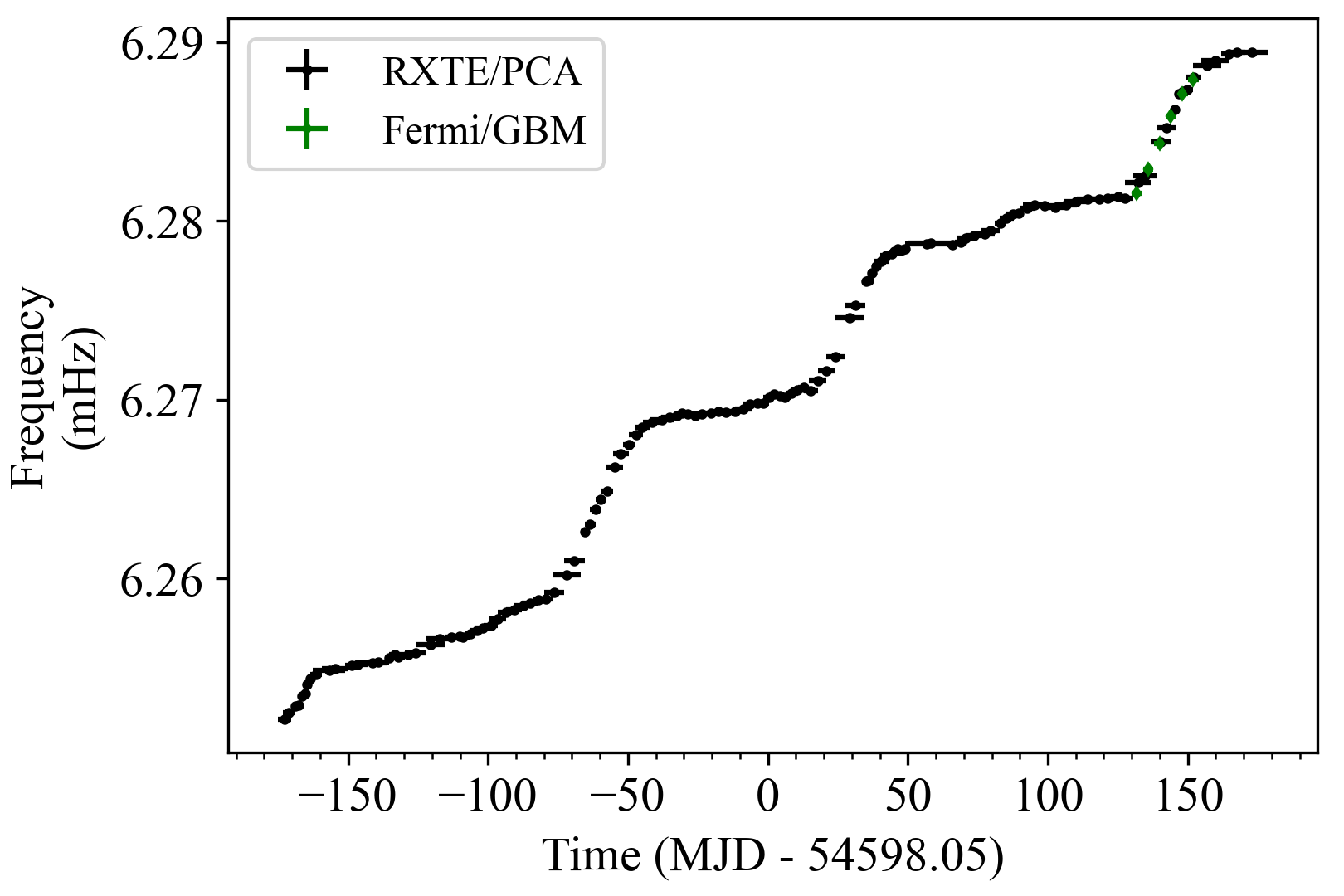}  \includegraphics[width=0.93\columnwidth]{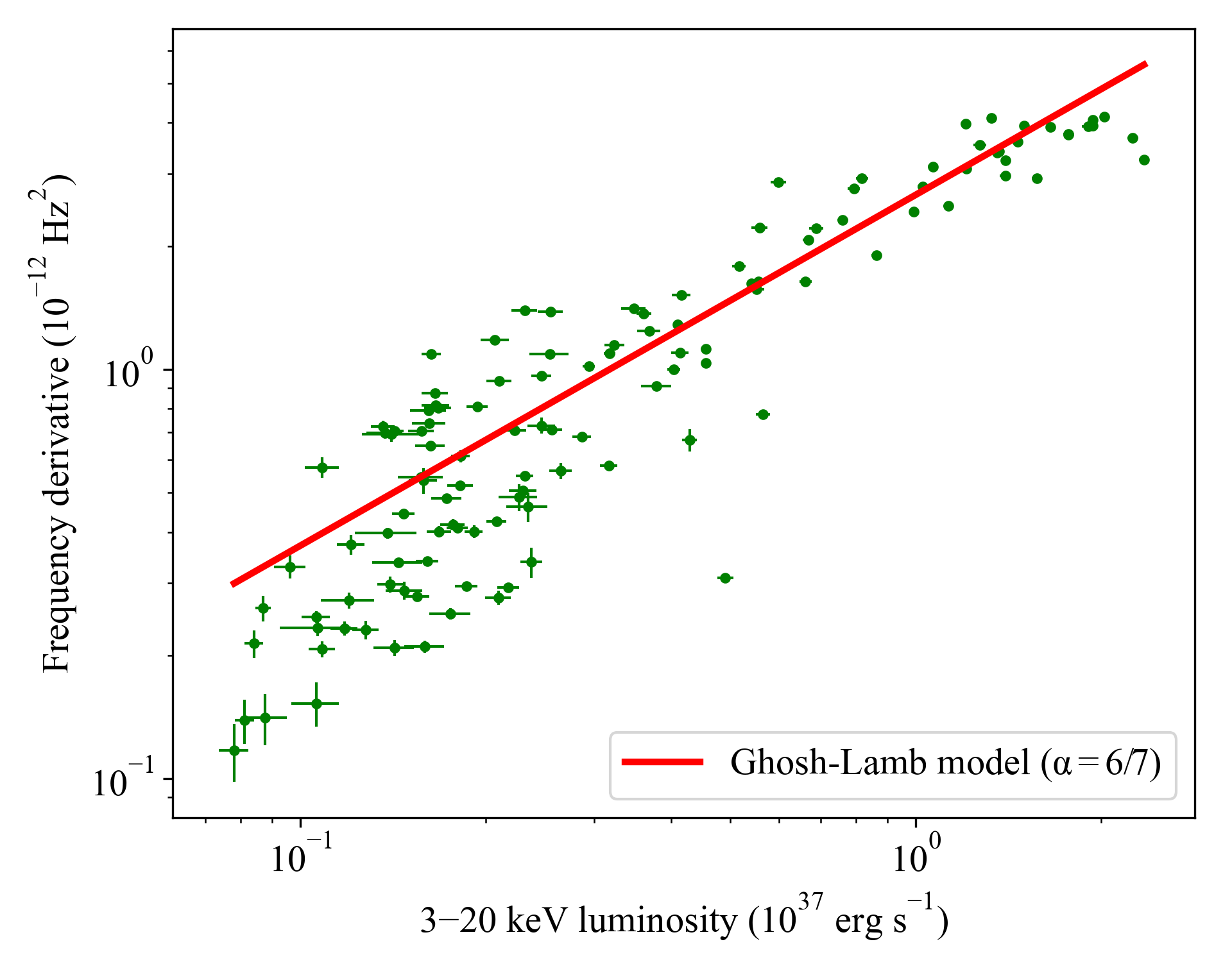}
    \caption{ Spin frequency history of MXB 0656--072 obtained with RXTE/PCA (black points, this work) and \emph{Fermi}/GBM (green diamonds) shown at the top. Lower panel gives the distribution of spin-up rate measurements as a function of the 3--20 keV RXTE/PCA luminosity of MXB 0656--072. The red line marks the best fit of the model.}
    \label{fig:f2}
\end{figure}

As the source continues to accrete, the pulse TOAs rapidly drift, especially due to enhanced accretion episodes at the type I outbursts within the analysis interval.
Therefore, we represent the pulse TOAs after the removal of the overall spin-up trend, $\Phi_{\mathrm{su}}$, starting at Epoch = MJD 54598.05, with $\nu=6.27128(6)$ mHz and $\dot{\nu}=1.14(2) \times 10^{-12}$ Hz s$^{-1}$. 
Noting the complex time evolution of the pulse TOAs even after elimination of the spin-up trend (Figure \ref{fig:f1}), we did not characterise them with high-order polynomials to specify a phase-coherent timing solution. Instead, we directly convert them into spin frequency measurements by fitting linear functions to every sequential TOA triplet.
The slope of these linear functions are utilised to calculate the frequency corrections at the middle times of TOA triplets by $\delta\nu = \delta\phi/\delta t$ over the spin-up trend mentioned above.
The resulting spin frequency history generated from the RXTE/PCA observations is illustrated in Figure \ref{fig:f2} (upper panel) together with the \emph{Fermi}/GBM measurements\footnote{\url{https://gammaray.nsstc.nasa.gov/gbm/science/pulsars/lightcurves/mxb0656.html}}.
The uncertainties of the frequency measurements are evaluated from the $1\sigma$ error ranges of the slope obtained during the fitting; and the horizontal uncertainties reflect the time range of TOA triplets contained therein.


\begin{figure}
\centering
        \includegraphics[width=1.0\columnwidth]{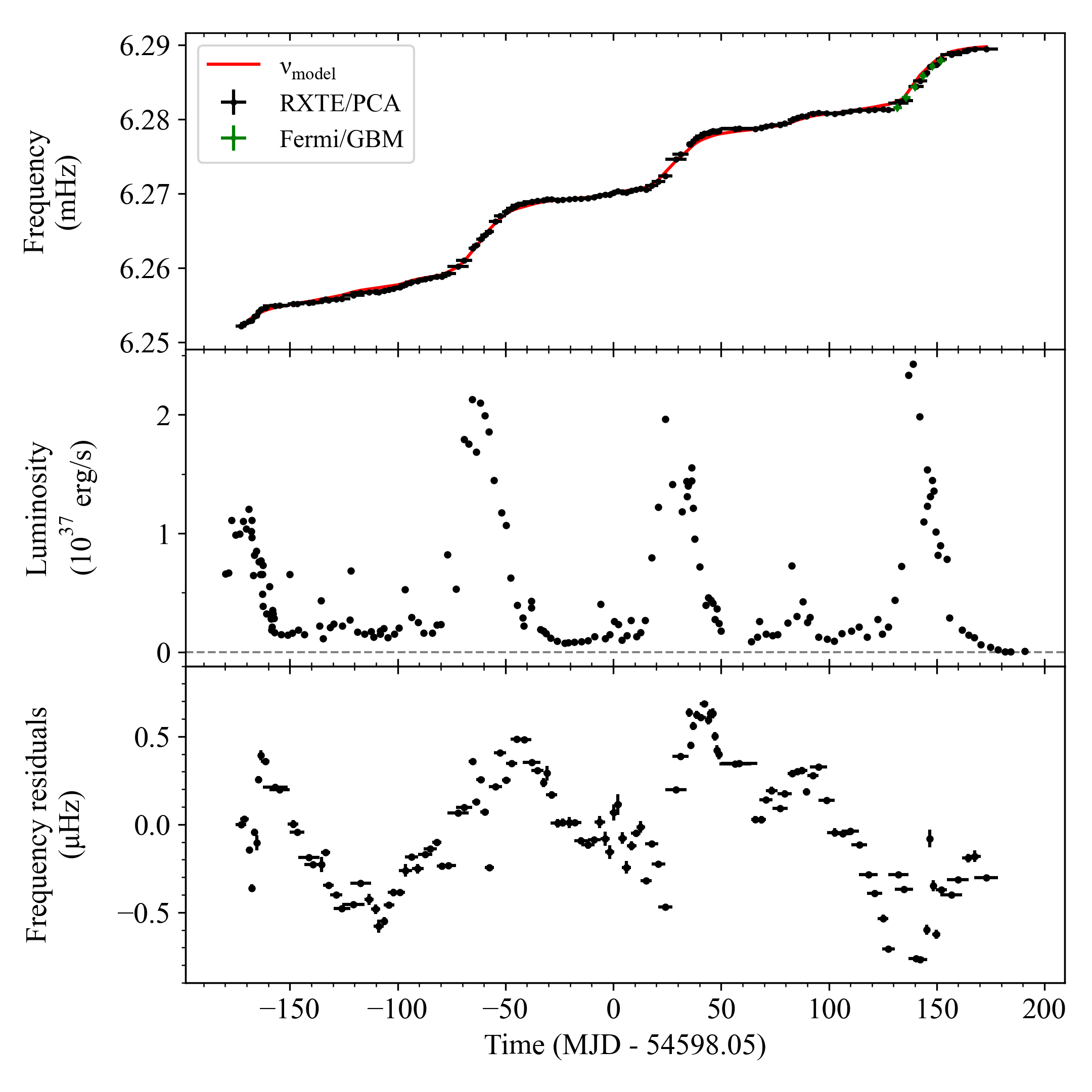}
    \caption{ Frequency measurements of MXB 0656--072 with RXTE/PCA (black) and \emph{Fermi}/GBM (green) shown at the top. The solid red line represents the intrinsic spin frequency model obtained through the $\dot{\nu}$--$L$ relation. Middle panel: 3--20 keV RXTE/PCA flux evolution. Lower panel: Spin frequency residuals after subtracting $\nu_{\textrm{model}}$ from the frequency measurements.}
    \label{fig:f3}
\end{figure}

As can be seen from Figure \ref{fig:f2}, MXB~0656--072 always spins up during the year, covered by the observations analysed in our study.
Even though the spin frequencies seem to modulate at the suggested orbital period \citep[$\sim$101.2 days;][]{2012Yan} over a general spin-up trend, they should be examined with caution. 
The spin-up rate enhancements correspond to the bursting episodes, yet the spin-up trend still persists at rather quiescent episodes between the type I outbursts.
Thus, we examine the intrinsic spin frequency evolution due to the torque exerted on the pulsar via accretion.
First, we calculated the spin frequency derivatives using sequential linear fits to $\sim$20-day-long segments in the frequency history in a similar procedure as described above.
Next, a linear spline interpolation of the RXTE/PCA flux evolution was used to estimate the flux range corresponding to the intervals where the spin frequency derivatives are measured.
The distribution of measured spin-up rates with respect to the flux exhibits a clear correlation (see Figure \ref{fig:f2}, lower panel). The spin-up rate increases with the luminosity, reaching up to $\sim$4$ \times 10^{-12}$ Hz s$^{-1}$ when the source achieves its maximum luminosity during this type I outburst sequence. It should be noted that the timing analysis of the type II outburst in 2003 revealed a time-averaged spin-up rate of $\sim$4.52(13)$ \times 10^{-12}$ Hz s$^{-1}$ \citep{2006McBride} for the average luminosity of $1.4(8)\times 10^{37}$ erg s$^{-1}$ (recalculated for 5.7 kpc), which is consistent with the results of our analysis.

Adopting the methodology specified in \cite{2004Galloway}, the correlation between the accretion torques and the X-ray luminosity is modelled in the form of a power-law profile, $\dot{\nu}\propto L_x^{\alpha}$, as expected from several torque models \citep[e.g.][]{1979Ghosh, 1995Lovelace, 2007Kluzniak}.
Fitting the relation with a power law yields an exponent $\alpha = 0.88\pm0.03$ (with 1$\sigma$ confidence level), which is in good agreement with the Ghosh–Lamb exponent, $\alpha = 6/7$.
It should be noted that 
MXB~0656--072 is a slow rotator ($P_{\mathrm{spin}}\sim 160$ s; $\omega_s << 1$); therefore, the effect of the curvature introduced by the fastness parameter $n(\omega)$ is insignificant in the given luminosity range. Furthermore, the observed source luminosities are mostly above $> 10^{36}$ erg s$^{-1}$ and different torque models \citep[e.g.][]{1995Wang,2009Enrico} does not yield notable changes in the correlation modeling and the spin frequency residuals.
 Therefore, we proceed our analysis using the prescription given in \cite{1979Ghosh} and freeze the exponent to 6/7 (i.e. $\dot{\nu}_{12}=\beta\, L^{6/7}_{37}$) during the modelling. 
In this case, the best fit yields $\beta=2.42(6)$. Then, the intrinsic spin frequency evolution originating from the accretion can be constructed following the approach outlined by \cite{2004Galloway} and \cite{2022Serim} as:
\begin{equation}
    \nu_{\mathrm{model}} (t) = \nu_0 + \int_{t_{0}}^{t_{f}} \beta L(t')^{6/7} dt'
    \label{eq:intrinsic}
,\end{equation}
where $t_{0}$ and $t_{f}$ denote the start and end times of the data set, and $\nu_0$ is the spin frequency at $t_0$. 
Figure \ref{fig:f3} displays the intrinsic spin frequency evolution obtained from this procedure in concert with the measured spin frequency set and its residuals.
The residuals after removal of the intrinsic spin frequency trend possess some structural variations; however, they do not  exhibit a clear sign for modulation with a profile that can be attributed to orbital motion. It is important to note that even when the exponent $\alpha$ is allowed to vary, its impact on residuals remains negligible. Therefore, the spin frequency residual profile is not well-suited for a fit to accurately determine a complete orbital solution.
Consequently, we are not able to constrain the orbital parameters of MXB~0656--072 with this data set. 

\section{Timing noise phenomenon}
\label{Sec4}
Timing noise is defined as the stochastic wandering in the residuals after fitting the data with an appropriate model.
There are different metrics to interpolate the timing noise amplitudes of pulsars \citep[e.g. see][]{1985Deeter,1994Arzoumanian,2010Shannon,2020Lower,2023Jones} and although there are numerous studies focusing on theoretical framework of underlying physical processes \citep[e.g. see][and references therein]{2021MeyersA,2021Meyersb,2023Antonelli,2023Vargas}, the timing noise phenomenon is still poorly understood.
In recent years, studies on timing noise have become especially important in the detection of gravitational waves \citep[e.g. see][and references therein]{2023Riles}.
Time-dependent noise characteristics of a pulsar provide hints for ongoing physical processes specific to that pulsar.
For example, in a binary system, timing noise at viscous timescales are dominated by the torque fluctuations associated with accretion flow \citep{2023SerimA}.
On the other hand, isolated pulsars are found to exhibit higher timing noise levels at lower characteristic ages and their timing noise behaviours are mostly associated with rotational instabilities and magnetospheric fluctuations \citep{2010Hobbs,2019CerriSerim,2022TJAA}.

The spin frequency history of MXB~0656--072 obtained in this work has a data sampling rate of 1 measurement per $\sim$3 days.
This is achieved through the comprehensive monitoring campaign conducted with RXTE/PCA, which provides an ideal input for timing noise analysis at moderate timescales. In order to understand the timing noise behaviour of the source at different timescales, three different techniques are employed.
The first technique utilises the root mean square (rms) values of the residuals after eliminating a polynomial of order $m$ from the input data spanning a time coverage of $T_{max}$ \citep{1972Boynton,1984Deeter,1985Cordes}.
In this case, the rms $\langle\sigma_r (m,T)\rangle$ values can be linked with a noise strength $S_r$ for an input data of duration, $T$, using the expression:
\begin{equation}
    S_r =\frac{\langle\sigma_r (m,T)\rangle}{\langle\sigma_r (m,1)\rangle_u}\frac{1}{T^{2r-1}}
,\end{equation}
where the subscript $r$ reflects the order of the red noise.
The coefficient $\langle\sigma_r (m,1)\rangle_u$ contains appropriate normalisation factors for the unit timing noise strength, $S_r$, on a unit timescale, $T=1$, which can be obtained via direct evaluations \citep[Table 1]{1984Deeter}.
Then, noise strength calculations were repeated for different timescales (i.e. $T_{max}/2^n$ for $n=1,2,3...$). Also, they were logarithmically combined across each timescale to acquire the corresponding power density estimate.
This technique enables the construction of a computationally less expensive low-resolution power density spectra (PDS) for torque fluctuations.
Afterwards, we further employed two new PDS generation techniques based on the temporal changes of the long-term noise strength estimators in the literature with the aim of improving the PDS resolution.

As a first new alternative technique, we employ the noise strength estimation method introduced by \cite{1994Arzoumanian}, which utilises the timing parameters obtained by fitting the input TOA or frequency set for a certain timescale.
In this case, the noise strength amplitudes are approximated using the spin frequency and its second order derivative as:
\begin{equation}
    \Delta (T) = \textrm{log} \bigg( \frac{|\ddot{\nu}|T^3}{6\nu}\bigg)
,\end{equation}
where $\ddot{\nu}$ represents the second derivative of the pulse frequency $\nu$.
In the literature, this method is used to calculate the noise strengths of many pulsars at the same timescale for comparison \citep[e.g.][]{1994Arzoumanian,2010Hobbs,2023Andersen,2023Zhou}.
The reference timescale for such comparisons is usually set to 10$^8$ s to examine long-term noise correlations \citep{1994Arzoumanian}; hence, the parameter is generally referred to as $\Delta_8$.
To trace the temporal noise behaviour of the source with this method, we use 10$^{\Delta}$ as the noise amplitude metric on the timescale, T.
The results are rescaled with a factor of $1.1\times10^{-19}$ s$^{-4}$ by cross-fitting the noise amplitudes obtained from the Deeter method, with the sole aim of comparing the PDS profiles.

To establish a second new alternative technique, we used a noise strength estimation metric known as the $\sigma_z$ method developed by \cite{1997Matsakis}.
With this method, the Allan variance of the residuals after the removal of the regular frequency evolution trend up to the $\ddot{\nu}$ term (with the polynomial description) is assumed to govern the timing noise. Then, $\sigma_z$ can be calculated as:
\begin{equation}
    \sigma_z = \frac{2}{\sqrt{5}} \frac{\langle\sigma_{\ddot{\nu}}\rangle}{\nu}
.\end{equation}
In this notation, $\langle\sigma_{\ddot{\nu}}\rangle$ has a correspondence to $\langle\sigma_1 (2,T)\rangle$ in the above description for the Deeter method, with a timescale-dependent renormalisation ratio of $2T\langle\sigma_r (m,1)\rangle_u / \sqrt{5}\nu$ for $\sigma_z$.
Similarly to $\Delta (T)$, $\sigma_z$ is also often used for investigating long-term correlations among large pulsar populations.

\begin{figure}
\centering
        \includegraphics[width=1.0\columnwidth]{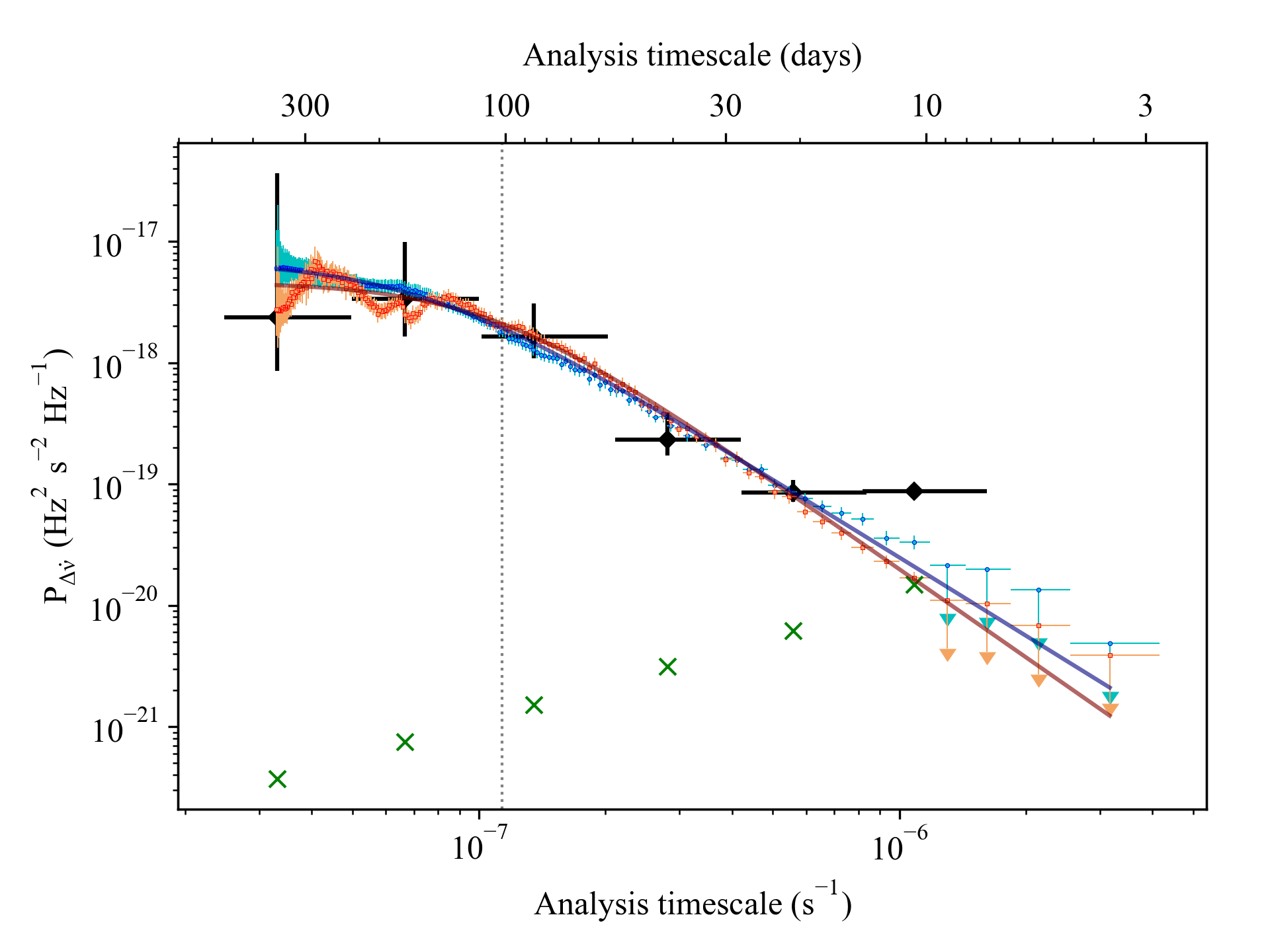}
    \caption{PDS of the torque fluctuations of MXB 0656--072 obtained via Deeter method (black dots), Delta-parameter variation (orange dots) and $\sigma_z$-variation (blue dots). Dark red and dark blue lines represent the best fit model for Delta-parameter and $\sigma_z$ variations using the model described through Equation \protect\ref{eq:noise}. The vertical dotted line indicates the orbital timescale suggested by \protect\cite{2012Yan}.}
    \label{fig:noise}
\end{figure}

To generate high-resolution PDSs for the latter two techniques, we calculated the noise strength amplitudes for temporal combinations in all of the possible timescales allowed by the input frequency data set.
To eliminate the possible outliers in the noise strength estimators, the estimators are filtered for valid fitting parameters.
The estimators undergo additional screening for measuremental noise strength levels that are associated with the uncertainties $\sigma_i$ of the input data set.
The measuremental noise strength level (MNS) in this case is calculated as:
\begin{equation}
   \textrm{MNS}(T) = \frac{\sum_i^N \sigma_i^2}{NT\langle\sigma_r (m,1)\rangle_u}
,\end{equation}
where $N$ is the total number of measurements and $\langle\sigma_r (m,1)\rangle_u$ is the normalisation factor described above. 
Then, the noise strength estimators are rebinned into power density estimators to construct the PDSs.
As suggested by \cite{1997Matsakis}, the uncertainty ranges ($\delta_{\pm}$) of the power density estimators can be obtained from $\chi^2$ distributions for $n$ degrees of freedom, which are specified by the number of noise estimations therein.
However, this error estimate is only applicable for independent non-overlapping segments at low $n$ \citep[see their Appendix A for details] {1997Matsakis}.
The uncertainty ranges of the overlapping segments with a large number of measurements are dominated by the standard error ($SE$) of the noise estimations.
Therefore, we used the combined error $\sqrt{\delta_{\pm}^2 + SE^2}$ to approximate the propagating uncertainties of the power density estimators.
Figure \ref{fig:noise} illustrates the PDSs resulting from the aforementioned three distinct techniques.

All the generated PDSs exhibit consistency across various timescales, with the exception of a slight difference at shorter timescales ($\lesssim$20 days).
In all cases, the presence of a red noise structure is clearly discernible, with a tendency to saturate at longer timescales ($\gtrsim$150 days).
This behaviour is reminiscent of observations in 2S 1417--624 \citep{2022Serim}, Her X-1, and Cen X-3 \citep{1997Bildsten,2023SerimA}.
The PDS continuum of MXB 0656--072 aligns with the generalised coluored exponential shot noise model proposed by \cite{1997Burderi}.
As such, we model the PDSs of torque fluctuations ($P_{\Delta\dot{\nu}}$), following the equation presented by \citep{1997Burderi,2023SerimA}:
\begin{equation}
\label{eq:noise}
P_{\Delta\dot{\nu}} = \frac{S_r}{1+ (\omega/\omega_{b})^\Gamma}.
\end{equation}
In this equation, $\omega$ represents the analysis frequency, $\omega_{b}$ stands for the break frequency that is induced by the decay timescale of exponential torque shots, while $\Gamma$ reflects the power law index that mediates the arbitrary slope triggered by the red noise process.
The results of the model fitting for all the PDSs are depicted in Figure \ref{fig:noise} and the corresponding parameters are provided in Table \ref{tab:noiseparam}.
Upon comparison, the latter two methods seem to provide an edge over the Deeter method, namely:
these methods yield PDSs with better resolution, particularly at higher frequencies.
The PDS modelling reveals a power law index of $\sim$-2, implying that the red noise continuum possibly originates from a monochromatic shot noise sequence rather than a coloured event sequence (i.e. torque events with varying timescales).

\begin{table}
\caption{Parameters obtained from PDS modelling through Equation \ref{eq:noise}.}
\centering
 \begin{tabular}{l c c c} 
 \hline
  &  $\Delta$-method  & $\sigma_z$-method & Unit  \\ 
       
  \hline      
  $S_r$    & $4.62(25) \times 10^{-18}$ & $7.02(13) \times 10^{-18}$ & Hz$^{2}$ s$^{-2}$ Hz$^{-1}$ \\
 $\omega_b$ & $1.05(7) \times 10^{-7}$ & $7.21(15) \times 10^{-8}$ & $s^{-1}$ \\
  $\Gamma$  & -2.42(35) & -2.15(6) &-\\
 \hline
 \end{tabular}
 \label{tab:noiseparam}
\end{table}

\section{Discussion and conclusion}
\label{Sec5}
MXB~0656--072 is a relatively understudied member of the Be/X-ray binary class.
We have revisited the type I outburst sequence that transpired during the period of 2007--2008, utilising observations from the RXTE/PCA instrument.
The methodology employed in our analysis involves pulse timing, which enables us to determine the temporal evolution of the pulse frequency for the specified time interval.
We discover a strong correlation between the spin frequency derivatives and source luminosity in the 3--20 keV band.
Given the assumption that in the 3--20 keV band, this luminosity  covers most of the bolometric luminosity in X-rays, the magnetic field strength can be estimated via torque models.
For example, this estimation can be done via the torque model introduced by \cite{1979Ghosh} as follows:
\begin{equation}
\label{eq:ghosh}
    \dot{\nu}_{12} = 1.4 \mu_{30}^{2/7} n(\omega_s)  R_6^{6/7} M_{1.4}^{-3/7} I_{45}^{-1} L_{37}^{6/7}
,\end{equation}
where $\dot{\nu}$ is the spin-up rate, $\mu$ is the dipole moment, $n(\omega_s)$ is the fastness parameter, and $L$ is the bolometric X-ray luminosity, while $R$, $M,$ and $I$  denote the radius, mass, and moment of inertia of the pulsar, respectively.
Our best-fit torque-luminosity relation is represented by the equation $\dot{\nu}_{12}=2.42(6) \times L_{37}^{6/7}$ (assuming a distance of 5.7 kpc, as specified in Section \ref{sec:1}).
By employing typical neutron star parameters and the slow rotator approximation ($n(\omega_s)\approx 1.4$), an estimation of the magnetic field strength of the source yields an approximate value of $4.2 \times 10^{12}$ G.
The calculated magnetic field strength is very close to (and even slightly higher than) the values inferred from CRSF  for MXB~0656–072, within the energy interval of 30–36 keV. This suggests the presence of a dipolar magnetic field with an estimated strength of approximately $B_{cyc} \approx 3.4 - 3.7 \times 10^{12}$~G \citep{2003Heindl, 2006McBrideProc, 2006McBride, 2012Yan}.

Theoretical investigations into the properties of accretion scheme at the surface of highly magnetised ($10^{11}-10^{13}$ G) NSs predict two distinct accretion regimes, known as subcritical and supercritical, in a given source where the transition between them is driven by the variations in accretion rate  \citep[e.g.][]{2012Becker,2015Mushtukov}. In the supercritical regime, the matter that flows in the accretion column towards the polar cap undergoes a radiation-dominated shock situated at a certain height above the neutron star's surface. In this regime, the radiation is expected to diffuse and escape through the column walls, which should lead to formation of a fan beam structure \citep{2012Becker}. In the subcritical accretion regime, matter flow is decelerated through Coulomb interactions before reaching to NS surface \citep{2021Mushtukov}. In this case, radiation propagates primarily vertical along polar caps which creates pencil beam structure. These states are also expected to render different spectral patterns as the typical emission height correlatively increases or decreases with luminosity in these different accretion regimes \citep[e.g.][]{2020Doroshenko,2022Serim,2023Rai,2023Mandal}.  
The onset of transition from subcritical to supercritical accretion regime is characterised by a critical luminosity via \citep{2012Becker}: \begin{equation}
\label{eq:becker}
    L_{\textrm{crit}} =1.5 \times 10^{37} B_{12}^{16/15}\,\, \textrm{erg s}^{-1}.
\end{equation}
where $B_{12}$ is the magnetic field strength of the pulsar in the units of $10^{12}$ G. Using Eq. \ref{eq:becker} and with the magnetic field strength derived from CRSF \citep{2003Heindl, 2006McBrideProc, 2006McBride, 2012Yan}, the critical luminosity falls within the range of $5.1-5.5 \times 10^{37}$ erg s$^{-1}$. It should be noted that \cite{2012Nespoli} reported an anti-correlation between the photon index and the luminosity, emphasising that the anti-correlation ceases  above a luminosity threshold of $\sim$2.3$\times 10^{37}$ erg s$^{-1}$, based on an assumed distance of 3.9 kpc. When adjusted for the \emph{Gaia} EDR3 distance of 5.7 kpc, this luminosity corresponds to approximately $\sim$4.9$\times 10^{37}$ erg s$^{-1}$.  
This luminosity threshold roughly agrees with the $L_{\textrm{crit}}$ calculated from $B_{cyc}$. However, it should be emphasised that the changes in the physical properties around the critical luminosity are also accompanied by the observational signatures in timing features, particularly on pulse profiles as the emission geometry switches between pencil and fan beam patterns \citep[e.g.][]{2017Epili,2018WilsonHodge,2020Tuo, 2022Serim,2022Wang}. The pulse profiles obtained in this work cover the X-ray luminosity below $\lesssim 3\times 10^{37}$ erg~s$^{-1}$, where the source accretes only in the subcritical regime with stable pencil beam emission. Further analyses of potential variations in pulse profiles around the critical luminosity are necessary to assess its validity.
\begin{figure}
\centering
\includegraphics[width=1.0\columnwidth]{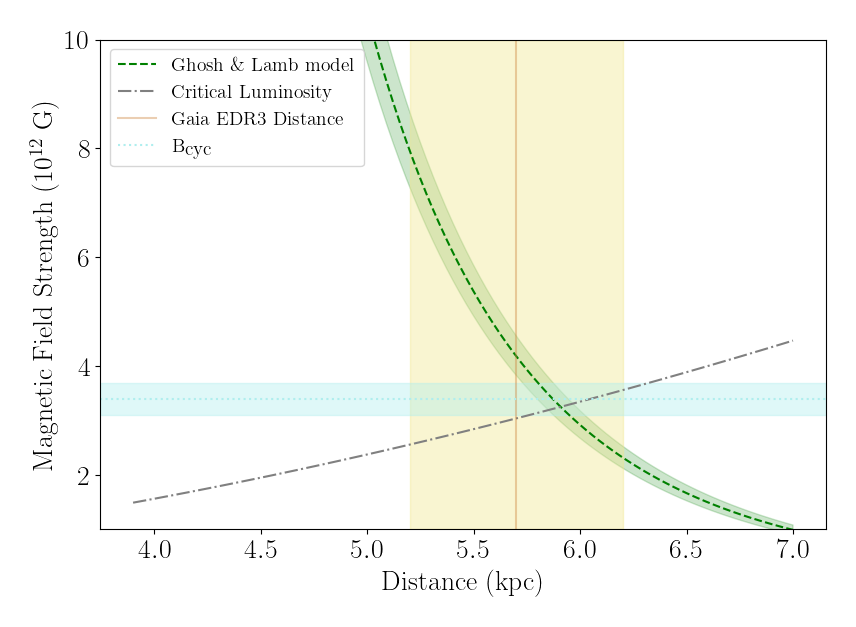}
    \caption{Magnetic field strength of MXB 0656--072 as a function of distance, deduced through various models. Shaded regions reflect the uncertainty range of measured and calculated parameters.}
    \label{fig:distance}
\end{figure}

On many occasions, torque models may point towards magnetic field strengths that significantly differ from the ones that are inferred from CRSFs \citep[see e.g.][for reviews]{2015Revnivtsev,2019Staubert}. Instead, for MXB~0656--072, they both point towards a magnetic field strength of the order of $\sim$$3-4\times 10^{12}$ G. To highlight the consistency of the results of different methods, we also present the distance (and, hence, the luminosity) dependence of the magnetic field strengths obtained from the \cite{1979Ghosh} and \cite{2012Becker} models utilising the spectral transition \citep{2012Nespoli} in Figure \ref{fig:distance}. This figure shows that the intersection of the magnetic field and distance obtained using the Ghosh \& Lamb model together with the \cite{2012Becker} model reside within the error box of new \emph{Gaia} distance estimate and $B_{cyc}$.

Furthermore, we constructed and examined the long-term spin frequency evolution of the source over an almost one-year period.
Overall, MXB~0656--072 undergoes spin-up during the type I outburst sequences that took place in 2007--2008.
The trend in spin frequency can be accurately depicted by the intrinsic frequency evolution driven by accretion torques.
Despite the structural variations in the residuals (see Figure \ref{fig:f3}, bottom panel), no distinct orbital modulation is evident.
Hence, we are unable to uncover the orbital timing parameters from this data set.
On the other hand, \cite{2012Yan} estimated the semimajor axis of the system as $253R_{\odot}$ by assuming an O9.7Ve companion of $20M_{\odot}$ and $15R_{\odot}$, considering its resemblance to A 0535+26.
Given these assumptions, the semi-major axis of the system can be converted to 587 lt-s.
Then, the maximum amplitude of frequency modulations $\delta\nu$ arising from orbital motion can be estimated as \citep{2017Serim}:
\begin{equation}
\label{eq:orbit}
\delta\nu  = \frac{2\pi}{P_{\textrm{orb}}}  \frac{1}{P_{\textrm{spin}}}  \frac{a}{c} \sin i 
,\end{equation}
where $P_{\textrm{orb}}$ is the orbital period, $P_{\textrm{spin}}$ is the spin period, $a/c$ represents the semi-major axis, and $\sin i$ denotes inclination. 
By applying Equation \eqref{eq:orbit}, the maximum amplitude of frequency modulations can be derived as $\sim$2.6 $\mu$Hz.
Nonetheless, the timing residuals in our analysis are constrained to an amplitude of $\sim$0.7 $\mu$Hz, suggesting that we are observing the system from a relatively `top view', with an inclination, $\sin i$, less than 0.27. This conclusion is consistent with deductions obtained from the detection of single-peaked Balmer emission lines in MXB 0656--072 \citep{2012Yan} and similar to the case of the Be/X-ray binary KS 1947+300 \citep{2023Liu}.

In addition, we performed an analysis of the PDS focusing on the variations in the spin frequency derivative of MXB~0656--072.
Our investigation is centred on a robust spin frequency data set with measurements obtained every $\sim$3 days, gathered from new measurements acquired through our timing analysis of RXTE/PCA observations.
Employing different methodologies, we generated three distinct PDS profiles to assess the temporal noise strength in the residuals.
While the initial method involves the conventional Deeter method \citep{1984Deeter}; the other two methods ($\sigma_z$: \citeauthor{1997Matsakis} \citeyear{1997Matsakis}, $\Delta$-parameter: \citeauthor{1994Arzoumanian} \citeyear{1994Arzoumanian}), to the best of our knowledge, are first to be applied for detailed PDS extraction.
The PDS profiles for MXB~0656--072 remain consistent across all these methods.

In a broader context, wind-fed sources exhibit a white noise pattern that is attributed to uncorrelated torque events produced by wind inhomogeneities.
Typically, such wind-fed systems exhibit torque noise strengths within the range of $10^{-20}$--$10^{-18}$ Hz$^2$ s$^{-2}$ Hz$^{-1}$ \citep{1997Bildsten}.
On the other hand, disc-fed sources rather exhibit a red noise pattern in their PDS in general, due to their torque evolution is being driven by a sequence of correlated events.
Such red noise patterns generally evolve with $\omega^{-2}$ and reach saturation at long timescales, potentially exceeding the viscous fluctuation timescales in the disc \citep{2011Icdem}.
Analysis of PDS in such persistent systems with disc accretion indicates that their torque noise strengths typically reside between $10^{-21}$--$10^{-18}$ Hz$^2$ s$^{-2}$ Hz$^{-1}$ \citep{1997Bildsten}.
The transient sources appear to share a similar nature in the PDS of torque fluctuations \citep{2007Baykal,2022Serim}, barring Swift~J0243.6+6124, which stands out with a relatively steep PDS ($\sim$$\omega^{-3.36}$) and a presence of a white noise component at shorter timescales \citep{2023SerimB}.
Yet, the pronounced steepness in the PDS of Swift J0243.6+6124 is argued to stem from quadrupolar interactions occurring at elevated luminosities \citep{2023SerimB}.
MXB~0656--072 exhibits a PDS profile akin to prior cases, such as Her~X-1, Cen~X-3 \citep{1997Bildsten,2023SerimA}, and 2S~1417--624 \citep{2022Serim}.
It is characterised by a red noise component with a steepness of $\Gamma \sim -2$ reaching saturation over a timescale of $\sim$150 days (see Table \ref{tab:noiseparam} for details).
The measured torque noise strength values of MXB~0656--072 fall within the range of $1 \times 10^{-20}$ to $7 \times 10^{-18}$ Hz$^2$ s$^{-2}$ Hz$^{-1}$ across the analysis frequency span of $3.5 \times 10^{-7}$ to $1 \times 10^{-6}$ s$^{-1}$.
This observed PDS profile is in line with the generalised coloured exponential shot model \citep{1997Burderi,2023SerimA}, which characterises an empirical model for the correlative sequence of events generated by accretion torques.
Since the steepness of the PDS is roughly consistent with $\Gamma\sim -2$, the profile suggests a likelihood of a monochromatic sequence of torque events generated by stable disc accretion throughout the observed time frame.

The newly implemented methods also handle estimations of temporal noise strength across all possible combinations of timescales within the input data, thereby providing improved PDS resolution, particularly at higher analysis frequencies.
Despite our sampling rate remains restricted to 1 measurement per $\sim$3 days for MXB~0656--072, in principle, these methods have the potential to distinguish properties within PDS at higher analysis frequencies.
Many studies in the literature focus on the theoretical aspects of the timing noise generation within a neutron star's inner response under the influence of external torques (e.g. \cite{1986Alpar,1991Baykal,1997Baykal,2017Gugercinoglu,2021MeyersA, 2023Antonelli}).
The expected timing noise contribution from the interior of a neutron star under such conditions is anticipated to be revealed at crust-core coupling timescales, approximately spanning from 10 to 100 rotational periods \citep{2009Sidery}.
Therefore, the techniques introduced in this paper could potentially enable the investigation of the PDS structure at higher analysis frequencies, thus providing an avenue to study neutron star interior through timing noise, given that the data sampling rate is adequate.

\begin{acknowledgements}
We thank the anonymous referee for his/her valuable remarks that improved this manuscript. Authors acknowledge the support from TÜBİTAK (The Scientific and Technological Research Council of Turkey) through the research project MFAG 118F037.
Danjela Serim was supported by the International Postdoctoral Research Fellowship Program (BİDEB 2219) of TÜBİTAK.
The authors thank Prof. Dr. Sıtkı Çağdaş İnam for his insightful comments.
Authors also thank Katja Pottschmidt, the Principal Investigator of the archival RXTE/PCA observation set P93032 used in this study.

The whole X-ray data used in this study is publicly available.
\emph{Fermi}/GBM frequency measurements are available at GBM Accreting Pulsars Project website (\url{https://gammaray.msfc.nasa.gov/gbm/science/pulsars.html}).

\end{acknowledgements}

%
%

\bibliographystyle{aa}
\bibliography{article}

\end{document}